# Tunable intrinsic ferromagnetic topological phases in bulk van der Waals crystal MnSb$_6$Te$_{10}$

Xin Zhang,[1,†] Shihao Zhang,[1,†] Zhicheng Jiang,[2,†] Shuchun Huan,[1] Yichen Yang,[2] Zhengtai Liu,[2] Song Yang,[3] Jinlong Jiao,[4] Wei Xia,[1,5] Xia Wang,[1,6] Na Yu,[1,6] Zhiqiang Zou,[1,6] Yongsheng Liu,[3] Jie Ma,[4] Dawei Shen,[2,7*] Jianpeng Liu,[1,5*] Yanfeng Guo[1,5*]

[1] School of Physical Science and Technology, ShanghaiTech University, Shanghai 201210, China

[2] State Key Laboratory of Functional Materials for Informatics, Shanghai Institute of Microsystem and Information Technology (SIMIT), Chinese Academy of Sciences, Shanghai 200050, China

[3] Shanghai Key Laboratory of Materials Protection and Advanced Materials in Electric Power, Shanghai University of Electric Power, Shanghai 200090, China

[4] Key Laboratory of Artificial Structures and Quantum Control (Ministry of Education), School of Physics and Astronomy, Shanghai Jiao Tong University, Shanghai 200240, China

[5] ShanghaiTech Laboratory for Topological Physics, ShanghaiTech University, Shanghai 201210, China

[6] Analytical Instrumentation Center, School of Physical Science and Technology, ShanghaiTech University, Shanghai 201210, China

[7] Center of Materials Science and Optoelectronics Engineering, University of Chinese Academy of Sciences, Beijing 100049, China

Intrinsic ferromagnetism is a crucial ingredient to realize quantum anomalous Hall effect in quasi two dimensional materials, thus the search of intrinsic ferromagnetic topological materials is one of the most concerned issues in the field of topological phases of matter. In this work, combining magnetotransport measurements, first principles calculations, and angle-resolved photoemission spectroscopy studies, we find that in MnSb$_6$Te$_{10}$, the n = 2 member of the MnSb$_2$Te$_4$/(Sb$_2$Te$_3$)$_n$ family, the strong magnetic competition realizes a fragile ferromagnetic ground state, which whereas easily enters into ferrimagnetic and the $\mathbb{Z}_2$ antiferromagnetic topological insulator phase with warming to higher temperature. Interestingly, the system stays in an inversion-symmetry-protected axion insulator phase in the ferromagnetic ground state as well as in the external magnetic field driven spin-polarized FM phase and can be converted into a Weyl semimetal with multiple Weyl nodes in the valence bands with hole doping, which are manifested by the measured notable intrinsic anomalous Hall effect. Our work thus provides an intrinsic magnetic topological material which is highly tunable into versatile topological phases by temperature, magnetic field, as well as carrier doping.

[†]The authors contributed equally to this work.

[*]Corresponding authors:
dwshen@mail.sim.ac.cn,
liujp@shanghaitech.edu.cn,
guoyf@shanghaitech.edu.cn.

**INTRODUCTION**

The magnetic topological materials provide extraordinary opportunities not only for clarifying the exact relation between magnetism and nontrivial topological electronic states, but also for discovering rich exotic unconventional topological quantum states such as the quantum anomalous Hall (QAH) insulators [1, 2], axion insulators [3-5] and magnetic Weyl semimetals (WSMs) [6-8]. It is widely acknowledged that when magnetically dope a topological insulator (TI), time-reversal ($\mathcal{T}$) symmetry would be broken and then an exchange gap would be opened in the gapless Dirac dispersion of the surface states [9, 10], which is capable of producing a chiral edge mode and hence the QAH effect [9, 11], i.e. the anomalous Hall conductance reaches an $e^2/h$ quantization plateau, when the Fermi level is exactly inside the surface exchange gap. Such case was observed in the Cr-doped $(Bi, Sb)_2Te_3$ thin film, in which the Cr doping breaks the $\mathcal{T}$ and opens a small Dirac mass gap (5 – 10 meV) in the topological surface state associated with the QAH effect through finely tuning the Fermi level into the gap [11-15]. Unfortunately, the magnetically doped TIs suffer the disorder and inhomogeneity caused by the randomly distributed dopants [16, 17], which inevitably renders the magnetization as non-mean-field-like and hence the realization of the QAH effect only in the sub-kelvin temperature range, far below its Curie temperature by orders of magnitude. However, this pioneering work naturally excited tremendous efforts to the exploration of intrinsic ferromagnetic (FM) topological phases that could realize the QAH effect at high temperature. In an array of intrinsic magnetic topological phases, such as the $Fe_3Sn_2$ [18, 19], $Mn_3Sn$ [20], $Co_3Sn_2S_2$ [21], the quantization is usually obstructed by the trivial bulk conduction channels at the chemical potential, making the observation of QAH impossible. In recent couple of years, the observation of QAH effect at ~ 4.5 K in the van der Waals (vdW) antiferromagnetic (AFM) insulator $MnBi_2Te_4$ has pushed this family forward as one of the research frontiers in topological physics [1, 2]. However, the in-plane ferromagnetically aligned $Mn^{2+}$ moments in $MnBi_2Te_4$ are antiferromagnetically coupled between adjacent "Te-Bi-Te-Mn-Te-Bi-Te" septuple layers (SLs), which requires the exfoliation of the crystal into few odd layers or very large external magnetic field to force the spins to be polarized into the FM state for the realization of QAH effect, thus largely hindering the wide usage of the exotic topological property. To realize an intrinsic FM coupling, one strategy is to insert nonmagnetic $Bi_2Te_3$ quintuple layer (QL) into the SLs to reduce the AFM coupling, which consequently facilitate the FM coupling in the peculiar A-type AFM structure with competing AFM and FM magnetic interactions. The $MnBi_2Te_4/(Bi_2Te_3)_n$ systems have therefore been subjected to intensive studies which show that $MnBi_8Te_{13}$, the n = 3 member in this family, realizes an intrinsic FM (Curie temperature $T_c$ below 10.5 K) TI with axion insulator state [3], while the QAHE has not been observed yet in this magnetic topological phase.

To realize the high temperature zero field dissipationless QAH effect, more intrinsic FM TIs as potential candidates are still extremely desired. However, construction of $MnBi_2Te_4/(Bi_2Te_3)_n$ with n > 3 is rather difficult and the present candidates are rather few, so the exploration of other analogues becomes extremely urgent. As the close relatives to the $MnBi_2Te_4/(Bi_2Te_3)_n$ family, the much less studied $MnSb_2Te_4/(Sb_2Te_3)_n$ family provides alternatives to realize the intrinsic FM TIs. The presently available members in the $MnSb_2Te_4/(Sb_2Te_3)_n$ family, $MnSb_2Te_4$ and $MnSb_4Te_7$ [22-27], unfortunately, are both with an AFM ground state and the realization of FM ground states through controlling the compositions can bring more disorder. The pristine $MnSb_2Te_4$ exhibits a topological phase from a topological trivial AFM insulator to an induced FM ideal WSM by the application of external magnetic field [22]. $MnSb_4Te_7$ in the AFM ground state is an axion insulator with gapped topological surface states, which can be converted into FM axion insulator by the application of small magnetic field [23]. Here, we report the discovery of the new member in the $MnSb_2Te_4/(Sb_2Te_3)_n$ family, the n = 2 member $MnSb_6Te_{10}$, seen in Figs. 1(a)-1(b), which hosts highly competing AFM and FM interactions that finally lead to a fragile FM ground state. First principles calculations indicate that MnSb$_6$Te$_{10}$ in the FM and AFM phases are both axion insulators, which are protected by inversion symmetry in the FM phase, and protected by the combination of time-reversal ($\mathcal{T}$) and half lattice translation symmetries ($\tau_{1/2}$) in the AFM phase, which are associated with gapped topological surface states. In the A-type AFM state, the interlayer AFM ordering can be converted into FM ordering under very small vertical magnetic field $\mu_0 H > 0.04$ T, and then MnSb$_6$Te$_{10}$ becomes a FM axion insulator protected by inversion symmetry. Moreover, the system in the FM phase would become a WSM upon slight hole doping, which is characterized by notable intrinsic anomalous Hall effect (AHE) caused by multiple Weyl nodes in the valence bands. Our work indicates that MnSb$_6$Te$_{10}$ is a promising magnetic topological material exhibiting both axion-insulator and Weyl-semimetal phases, which can be tuned flexibly by temperature, magnetic field and carrier doping.

The details for crystal growth, compositions characterization, magnetization, magnetotransport, ARPES measurements, and first principles calculations of the bulk MnSb$_6$Te$_{10}$ are presented in Supplementary Information (SI) which includes references [22, 23, 28-32, 3, 33-39].

**RESULTS AND DISCUSSION**

A picture of a typical MnSb$_6$Te$_{10}$ single crystal is shown in the inset of Fig. 1(b), with the size of about 2 ×2 ×0.3 mm$^3$ and the flat surface corresponding to the *ab* plane. Fig. 1(b) shows the room temperature powder X-ray Bragg reflections from the *ab* plane of the single crystal, which can be nicely indexed by using the MnBi$_6$Te$_{10}$ structure model [40, 41] without other impurity peaks. It should be noted that we used a square root scale to plot the peak intensity. The MnSb$_6$Te$_{10}$ crystallizes in the $R\bar{3}m$ (No. 166) space group with the lattice parameters $a = b = 4.236$ Å, $c = 101.25$ Å, $\alpha = \beta = 90^o$, $\gamma = 120^o$. The *c* axis (101.25 Å) is apparently longer than that (23.265 Å) of

MnSb$_4$Te$_7$ [23].

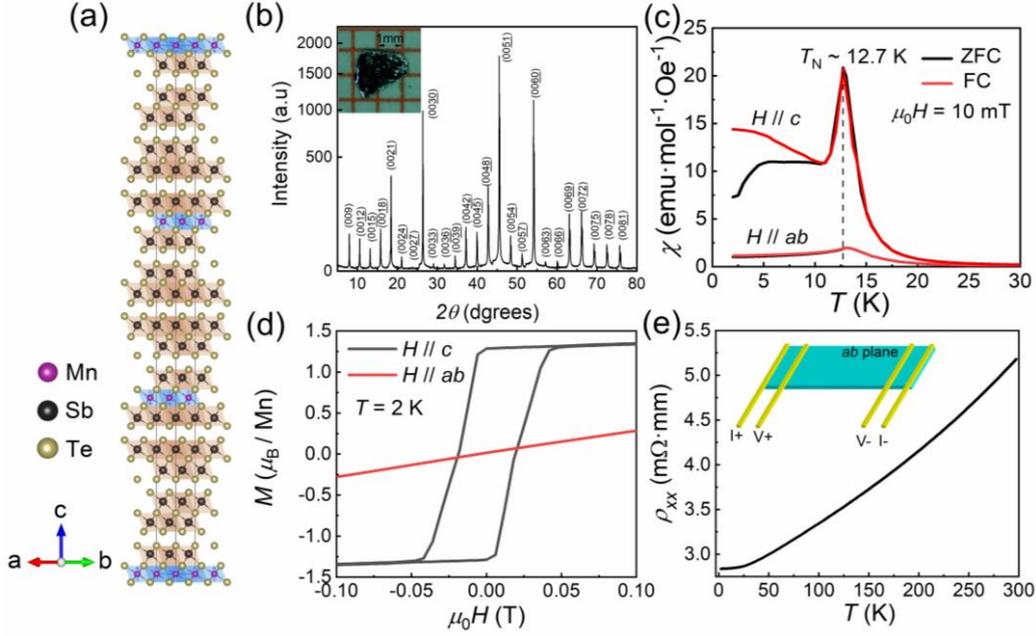

**Fig. 1.** (a) The crystal structure of MnSb$_6$Te$_{10}$ with space group $R\bar{3}m$. The blue arrows represent the Mn$^{2+}$ spins in the A-type AFM structure. Brown block: edge-sharing SbTe$_6$ octahedra. Blue block: edge-sharing MnTe$_6$ octahedra. (b) The indexed room temperature powder X-ray diffraction peaks from the *ab* plane of MnSb$_6$Te$_{10}$ crystal in square root scale. Insert: Image of a typical MnSb$_6$Te$_{10}$ single crystal synthesized in this work. (c) The temperature dependence of susceptibility χ under $\mu_0 H$ = 10 mT for *H//c* and *H//ab* plane. (d) Isothermal magnetizations at 2 K for *H//c* and *H//ab* plane. (e) The temperature dependent of transverse resistivity $\rho_{xx}$ at $\mu_0 H$ = 0 T measured from 2 K to 300 K. The sketch for the four-probe measurement configuration is inserted.

The resistivity and magnetization of MnSb$_6$Te$_{10}$ are presented in Figs. 1(c)-1(e), respectively. Seen in Fig. 1(c), the zero-field-cooling (ZFC) and field-cooling (FC) magnetic susceptibility (χ) measured at $\mu_0 H$ = 0.01 T with *H//c* axis and *H//ab* plane, respectively, displays an abrupt transition around $T_N$ = 12.7 K, signifying the intrinsic long-range AFM ordering, which is slightly lower than that of MnSb$_2$Te$_4$ (~ 19 K) and MnSb$_4$Te$_7$ (~ 13.5 K) [22, 23]. The χ shown in Fig. 1(c) decreases fast for *H//c* while gently for $H//ab$, exposing an anisotropic AFM exchange with the $c$ axis being the magnetic easy axis. The $\chi$ below $T_N$ for $H//c$ exhibits a pronounced bifurcation below 10 K, suggesting the formation of FM domains. It is also notable that the magnitude of $\chi$ with $H//c$ is much larger than that for $MnSb_2Te_4$ and $MnSb_4Te_7$, even by one order of magnitude, likely indicative of different ground states. This hypothesis is supported by the isothermal magnetization ($M(H)$) data measured at 2 K [Fig. 1(d)] as well as results of the Curie-Weiss plot to the magnetic susceptibility as presented in the SI by Fig. S1. For $H//ab$ plane, the $M(H)$ is basically linear and decreases gently with further cooling, while that for $H//c$ axis shows a clear hysteresis loop without appreciable multi-step features caused by the spin-flop transitions that are usually observed in AFM $MnBi_2Te_4/(Bi_2Te_3)_n$ and $MnSb_2Te_4/(Sb_2Te_3)_n$, thus ambiguously demonstrating a FM ground state in $MnSb_6Te_{10}$. Moreover, the Curie-Weiss plot gives the Weiss temperature of about ~ 11 K, also supporting the FM ground state. However, as seen in Fig. S2, the FM ground state is rather fragile, which is easily converted into a ferrimagnetic state with warming to a higher temperature, indicated by the clear multiple-step features in the isothermal magnetizations at temperatures larger than 2 K. Our calculations reveal that the total energy difference between the AFM and FMz states is only of 116 $\mu$eV per Mn atom pair, which is much smaller than that of $MnSb_2Te_4$ and $MnSb_4Te_7$ system and roughly consistent with that presented earlier [42]. Despite of the AFM ground state proposed by both calculations of ours and that in Ref. 42, the magnetization chracterizations support the FM ground state for $MnSb_6Te_{10}$. The coercivity of $MnSb_6Te_{10}$ is ~ 50 mT and the effective moment $\mu_{eff}$ = 4.54$\mu_B$/Mn, which is close to the expected 5$\mu_B$/Mn for $Mn^{2+}$. The weak interlayer AFM coupling can be easily converted into FM order by vertical magnetic field, suggested by the saturation of magnetic moment at larger magnetic field at around 0.04 T. The temperature dependence of longitudinal resistivity $\rho_{xx}$ at $\mu_0 H = 0$ T with electrical current $I//ab$ plane is presented in Fig. 1(e), which shows a monotonic decrease upon cooling down to 2 K and does not appreciably exhibits anomaly corresponding to the AFM transition, possibly due to the weaker magnetic fluctuations and the limited scattering strength on carriers.

The bulk MnSb$_6$Te$_{10}$ crystal has $P$, $M$, $C_{3z}$, and $C_2'$ symmetries without any magnetic order. The AFM state breaks the $M$ mirror symmetry and $C_2'$ rotational symmetry, but preserves $P$ inversion symmetry and an additional nonsymmorphic symmetry which is the combined symmetry of time-reversal symmetry ($\mathcal{T}$) followed by $\tau_{1/2}$ sliding. Besides the rather small total energy difference between the AFM and FMz states, our calculations also show that, if the system enters into the FM state, the total energy of FMx state is only 2 $\mu$eV lower than that of FMz state, while 0.4 meV lower than that of FMy state. Thus the competition between different spin-polarized FM phases will be stronger as compared with MnSb$_2$Te$_4$ and MnSb$_4$Te$_7$ systems.

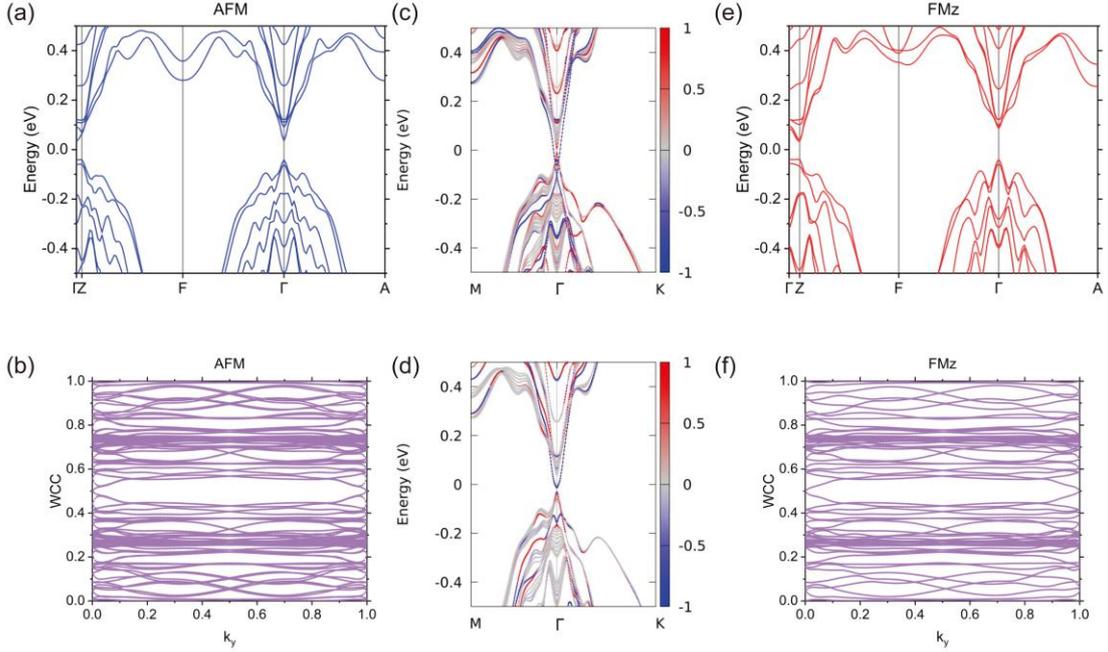

**FIG. 2.** The calculated electronic band structure for MnSb$_6$Te$_{10}$. (a) The band structures of antiferromagnetic (AFM) state. Here the k-point path is along $\Gamma$(0,0,0) – Z(0,0,1/2) – F(1/2,0,1/2) – $\Gamma$(0,0,0) – A(1/2,0,0). (b) The Wannier charge center (WCC) evolution in AFM state at $k_z = 0$ plane. Here the surface states are remarked as red line (stands for SL–QL–QL... surface) or blue

line (stands for QL–QL–SL... surface) in fig(c). But in (d), two types of surface are both QL–SL–QL... surface. The grey lines refer to bulk state. The energy bands of ferromagnetic state (e) and WCC evolution of this magnetic configure (f) are also present.

Now we turn to the electronic structure of $MnSb_6Te_{10}$. The energy bands of this system in different magnetic states show a direct bandgap of ~ 80 meV. The Wilson loops presented in Figs. 2(b) and 2(f) indicate nontrivial topological states in the system. Because the $MnSb_6Te_{10}$ system has inversion symmetry, we can choose the symmetry indicator $Z_4 = \sum_{k=1}^{8}(n_k^+ - n_k^-)/2$ mod 4 to characterize the topological characters of the different magnetic topological phases where $n_k^+/n_k^-$ is the number of occupied states with even or odd parity at the inversion-invariant momenta $k$ [43-46]. Here $Z_4 = 2$ at different magnetic states means that the system is a robust axion insulator under different magnetic configurations, which is similar as $MnSb_4Te_7$. Specially, the combined symmetry $\mathcal{T}\tau_{1/2}$ preserves the AFM state as an AFM TI. In this topological platform, the behavior of surface states depends on the types of specific surface. There are three kinds of surface: SL–QL–QL... A-type surface, QL–QL–SL... B-type surface and QL–SL–QL... C-type surface. It is noted that the combined symmetry $\mathcal{T}\tau_{1/2}$ is broken in these three types of surfaces. The A-type surface is terminated with SL layer and thus has gapped Dirac band dispersion whose gap originates from Zeeman coupling to the surface magnetizations. In the B-type surface, the large hybridization between the states in the QLs and SLs creates the surface states throughout the bulk gap and makes the system metallic at the surface. As for the C-type surface, the hybridization is weakened as compared with the B-type surface and therefore still produces a gapped surface state.

We continue to discuss the magnetotransport properties of $MnSb_6Te_{10}$ in different magnetic configurations. The synthesized bulk $MnSb_6Te_{10}$ crystal is apparently hole doped and behaves as a metal within the measured temperature range, seen in both Fig. 1(e) and Fig. S2. Fig. 3(a) shows the magnetic field dependence of longitudinal resistivity $\rho_{xx}(H)$. It is seen that when the temperature is larger than 2 K, the AFM state is manifested by the butterfly shaped $\rho_{xx}(H)$ at weak magnetic fields $H < H_C$ ($\mu_0 H_C \sim 50$ mT) at temperatures lower than $T_N$. The critical field of $\rho_{xx}(H)$ is well consistent with the hysteresis loop seen in the $M(H)$ and changes dramatically during the spin-flop process. When the magnetic field is further increased, $\rho_{xx}(H)$ exhibits a wavy shape trend which may come from the contribution of the Sb$_2$Te$_3$ layer. The anomalous Hall resistivity $\rho_{xy}^A$ obtained after subtracting the normal Hall resistivity $\rho_{xy}^N$ ($= R_0\mu_0 H$) from $\rho_{xy}$ exhibits a notable hysteresis loops as shown in Fig. 3(b), which highly resembles the $M(H)$ curve at 2 K as seen in Fig. 1(d) and Fig. S3. The magnitude and the area of the hysteresis loops in $\rho_{xy}^A$ all gradually diminish with the increase of temperature for $H > H_C$ and $T < T_N$, and they eventually almost vanish when temperature is higher than $T_N$. Importantly, unlike $MnSb_2Te_4$ and $MnSb_4Te_7$, the saturation behavior of $\rho_{xy}^A$ can be visible at 2 K even without external magnetic field, ambiguously unveiling a zero-field AHE in $MnSb_6Te_{10}$. This different behavior highlights the realization of the FM ground state in $MnSb_6Te_{10}$.

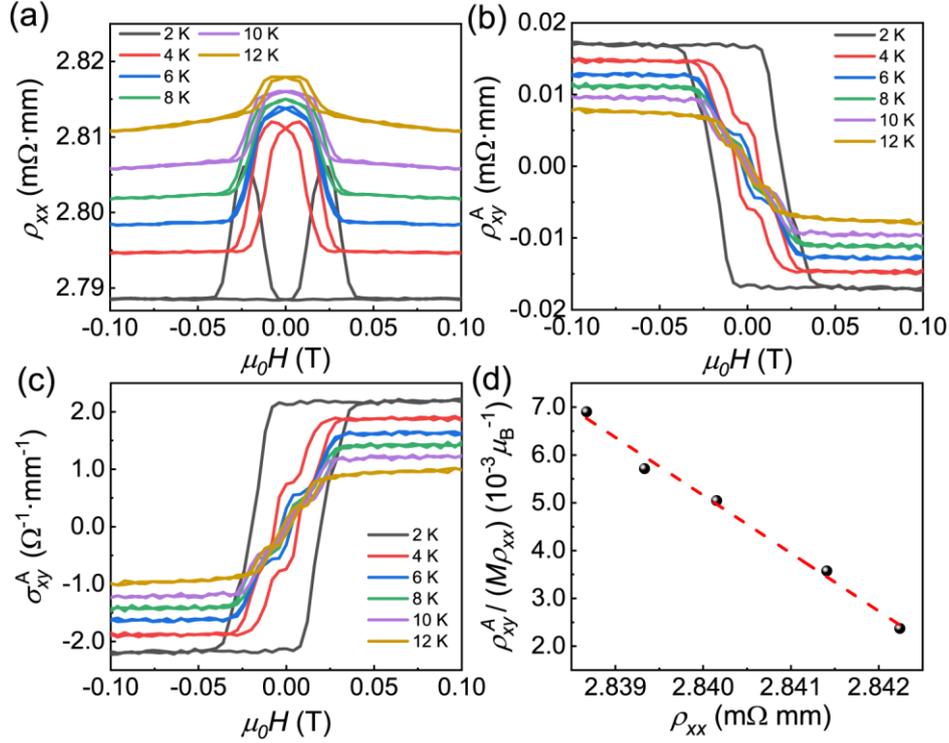

**FIG. 3.** Magnetotransport properties of bulk MnSb$_6$Te$_{10}$ single crystal. (a) Magnetoresistance with $I//ab$ plane and $H//c$ axis at various temperatures. (b) The anomalous hall resistivity $\rho_{xy}^A$. (c)The anomalous Hall conductivity $\sigma_{xy}^A$ at various temperatures. (d) The plot of $\rho_{xy}^A/(M\rho_{xx})$ vs. $\rho_{xx}$ where the red dashed line denotes the linear fit when temperature below 12 K.

The converted anomalous Hall conductivity (AHC) $\sigma_{xy}^A$ ($=\rho_{xy}^A/(\rho_{xy}^{A\,2}+\rho_{xx}^2)$) is presented in Fig. 3(c). In a magnetic topological system, it is necessary to trace the real origin for the AHE since other extrinsic factors such as skew and side-jump contributions besides the Berry curvature could also contribute to the AHE [47]. We therefore used the so-called TYJ scaling method to determine the dominant mechanism for the AHE [48], which has been widely used for the analysis for AHE effect in magnetic topological phases [49-52]. Within the framework of TYJ scaling method, the total Hall resistivity could be expressed as $\rho_{xy} = \rho_{xy}^N + \rho_{xy}^A = R_0 B + R_s 4\pi M$, where $R_0$ is the normal Hall coefficient, and $R_s$ is the anomalous Hall coefficient. A more specific formula of $\rho_{xy}^A$ including longitudinal resistivity is $\rho_{xy}^A = a(M)\rho_{xx} + b(M)\rho_{xx}^2$, where the first term denotes the extrinsic contributions including the skew component, while the second term represents the intrinsic contributions which also include the side-jump component [53]. Therefore, according to the above formula, $\frac{\rho_{xy}^A}{M\rho_{xx}} = b\rho_{xx}$ is from intrinsic contribution from the nontrivial topological states. Seen in Fig. 3(d), the relation between $\rho_{xy}^A/(M\rho_{xx})$ and $\rho_{xx}$ is linearly when the temperature is below 12 K, suggesting intrinsic AHC in $MnSb_6Te_{10}$. At 2 K, $\rho_{xy}^A$ is extracted to be 17 Ω mm and $\sigma_{xy}^A$ is 2.2 Ω$^{-1}$ mm$^{-1}$, which are rather close to the theoretical values as discussed in the SI.

To determine the nontrivial topological properties of $MnSb_6Te_{10}$ experimentally, we used synchrotron based high-resolution angle resolution photoemission spectrum (ARPES) to study its low-energy band structures along the high-symmetry K-Γ-K direction at 20 K, i.e. in the paramagnetic region, as shown in Fig. 4. Systematic photon energy dependent measurements are capable of distinguishing the surface and bulk bands states near the Fermi level, as shown in Fig. 4(a). By comparing with the first principles calculations results of K-Γ-K direction shown in Fig. 4(d), we can recognize the periodically changed bulk bands indicated by the green arrows and invariable surface bands marked by the red arrows. The intensity of surface bands is suppressed by the bulk bands at some specific photon energies. As illustrated in Figs. 4(b) and 4(c), the shoulder-like bands located at 100 meV below the Fermi level taken with 60 eV are relatively weaker as compared with those taken at 75 eV. We thus chose the 70 eV photons to study the band structure near the Fermi level, where the intensities of surface and bulk bands are comparable. Figures 4(d) and 4(e) present the raw and second derivative ARPES E-k maps, respectively. Remarkably, $MnSb_6Te_{10}$ shows similar p-type carrier dosage as $MnSb_4Te_7$. By a detailed comparison with the calculations presented in Fig. 4(f), we can evaluate that the Fermi level of the as-grown $MnSb_6Te_{10}$ is located at approximately 380 meV below the Dirac point. The topological protected surface states marked by the red arrows can be well distinguished through the second derivative spectrum. $MnSb_6Te_{10}$ is supposed to own three types of cleave planes, but the band structured of them are indistinguishable near the Fermi level as shown in Fig. 2. To further resolve these Dirac surface states, we tried to tune its chemical potential by Bi substitution for Sb in $MnSb_6Te_{10}$ as well as by alkali metal doping. As shown in Fig. S6, with 20% Bi substitution, the $E_F$ of $Mn(Bi_{0.2}Sb_{0.8})_6Te_{10}$ shifts upward by approximately 40 meV. At the same time, the multi bulk bands marked by green arrows in Figs. 4(e) and 4(f) become much clearer. Therefore, by ARPES measurement, the existence of non-trivial topological surface states is confirmed in $MnSb_6Te_{10}$.

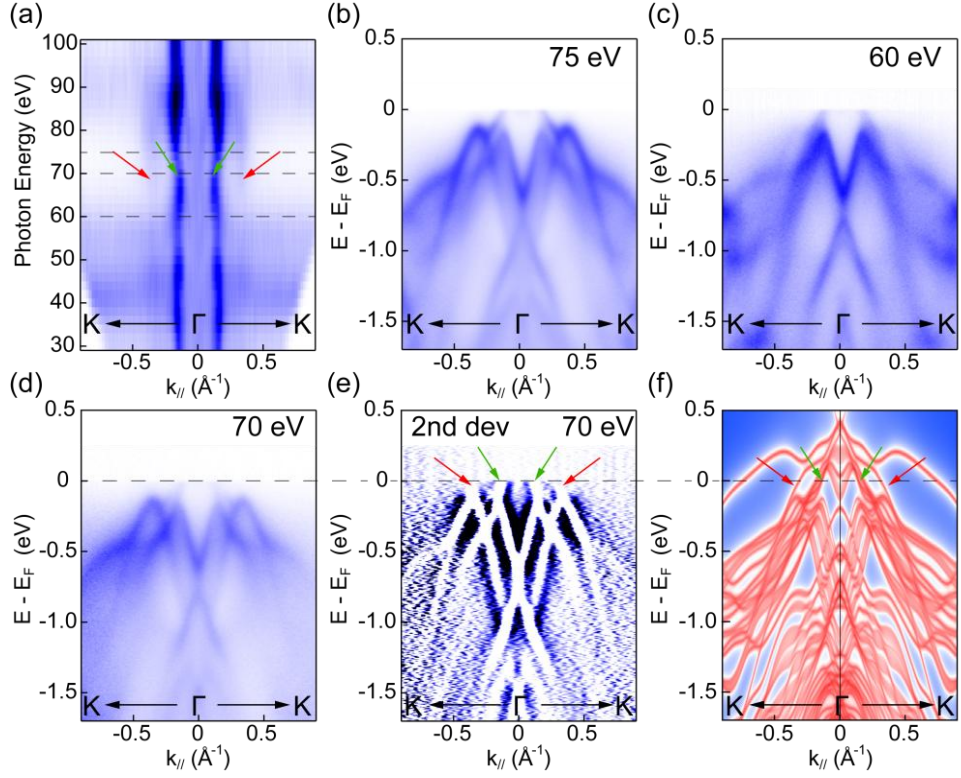

**FIG. 4.** ARPES measurement results of MnSb$_6$Te$_{10}$. (a) Photon energy dependent ARPES mapping alone K-Γ-K direction. (b), (c) ARPES k$_{//}$-E intensity spectrum along K-Γ-K collected at 75 and 60 eV, respectively. (d), (e) ARPES intensity and its second derivative spectrum along the K-Γ-K direction collected at 70 eV, respectively. (d) Calculated band structure on the SL-QL-QL… termination.

**SUMMARY**

To summarize, in this work we report the first successful growth of high-quality $MnSb_6Te_{10}$ single crystal, and have performed systematic magnetotransport measurements, first principles calculations, and ARPES studies on its magnetic and topological properties. Unlike other already known $MnSb_2Te_4/(Sb_2Te_3)_n$ (n = 0 and 1), we find that bulk $MnSb_6Te_{10}$ has a fragile ferromagnetic ground state. We also deduce that bulk $MnSb_6Te_{10}$ would stay in axion insulator states for both FM and AFM configurations, and the topologically nontrivial states are confirmed by our first principles calculations and ARPES measurements. The ferromagnetic $MnSb_6Te_{10}$ thus realizes a natural ferromagnetic axion insulator state. Moreover, there are multiple Weyl nodes in the FM phase such that the system behaves as a Weyl semimetal upon carrier doping, which is characterized by notable anomalous Hall effect. Therefore, the $MnSb_6Te_{10}$ in the FM configuration realizes a unique phase with co-existing axion-insulator and Weyl-semimetal states, and the transition between the two topological states can be tuned by carrier doping. Our work is a significant step forward in the synthesis of $MnSb_2Te_4/(Sb_2Te_3)_n$ family magnetic topological phases as well as in understanding the intriguing physical properties of magnetic topological materials.

**ACKNOWLEDEMENTS**

The authors acknowledge the support by the National Natural Science Foundation of China (Grant Nos. No. 92065201, 11874264 and U2032208), the National Key R&D Program of the MOST of China (Grant No. 2016YFA0300204) and the Shanghai Science and Technology Innovation Action Plan (Grant No. 21JC1402000). S. Zhang and J. Liu acknowledge support from National Science Foundation of China (grant no. 12174257) and the start-up grant of ShanghaiTech University. Y. F. Guo acknowledges the start-up grant of ShanghaiTech University and the Program for Professor of Special Appointment (Shanghai Eastern Scholar). Part of this research used Beamline 03U of the Shanghai Synchrotron Radiation Facility, which is supported by ME2 project under contract No. 11227902 from National Natural Science Foundation of China. The authors also thank the support from the Analytical Instrumentation Center (SPST-AIC10112914), SPST, ShanghaiTech University.

# Supplementary Information

# Tunable intrinsic ferromagnetic topological phases in bulk van der Waals crystal MnSb$_6$Te$_{10}$

Xin Zhang,[1,†] Shihao Zhang,[1,†] Zhicheng Jiang,[2,†] Shuchun Huan,[1] Yichen Yang,[2] Zhengtai Liu,[2] Song Yang[3], Jinlong Jiao[4], Wei Xia[1,5], Xia Wang[1,6], Na Yu[1,6], Zhiqiang Zou,[1,6] Yongsheng Liu,[3] Jie Ma,[4] Dawei Shen,[2,7*] Jianpeng Liu[1,5*], Yanfeng Guo[1,5*]

[1] School of Physical Science and Technology, ShanghaiTech University, Shanghai 201210, China

[2] State Key Laboratory of Functional Materials for Informatics, Shanghai Institute of Microsystem and Information Technology (SIMIT), Chinese Academy of Sciences, Shanghai 200050, China

[3] Shanghai Key Laboratory of Materials Protection and Advanced Materials in Electric Power, Shanghai University of Electric Power, Shanghai 200090, China

[4] Key Laboratory of Artificial Structures and Quantum Control (Ministry of Education), School of Physics and Astronomy, Shanghai Jiao Tong University, Shanghai 200240, China

[5] ShanghaiTech Laboratory for Topological Physics, ShanghaiTech University, Shanghai 201210, China

[6] Analytical Instrumentation Center, School of Physical Science and Technology, ShanghaiTech University, Shanghai 201210, China

[7] Center of Materials Science and Optoelectronics Engineering, University of Chinese Academy of Sciences, Beijing 100049, China

[†]These authors contributed equally to this work.

*Corresponding authors:

dwshen@mail.sim.ac.cn,

liujp@shanghaitech.edu.cn,

guoyf@shanghaitech.edu.cn.

### a. Crystal growth and characterizations

The MnSb$_6$Te$_{10}$ single crystals were grown by using the self-flux method. Starting materials of Mn (99.95%), Sb (99.999%) and Te (99.999%) blocks were mixed in a molar ratio of 1: 10: 16 and placed into an alumina crucible which was then sealed into a quartz tube in vacuum. The assembly was heated in a furnace up to 750 °C within 10 hrs, kept at the temperature for 15 hrs, then cooled down to 625 °C at a temperature decreasing rate of 10 °C/hr and slowly cooled down to 612 °C. The excess melt components were removed at this temperature by quickly placing the assembly into a high-speed centrifuge and black crystals with shining surface in a typical dimension of 2 × 2 × 0.3 mm$^3$, shown by the insert picture as Fig. 1(b) in the main manuscript.

The phase and quality of the single crystals used in this work were examined on a Bruker D8 Advance powder X-ray diffractometer (PXRD) with Cu $K_{\alpha 1}$ ($\lambda$ = 1.54184 Å) at 298 K.

### b. Magnetotransport measurements

Magnetic properties of MnSb$_6$Te$_{10}$ were characterized on a commercial magnetic property measurement system. The temperature dependent magnetic susceptibility ($\chi(T)$) along the in-plane ($H//ab$) direction was measured in the zero-field cooling (ZFC) and field-cooling (FC) mode in the temperature range of 1.8 to 350 K. The temperature dependent isothermal magnetizations ($M(H)$) were measured within the temperature range of 2 - 15 K and magnetic field range of -3 - 3 T. Magnetotransport measurements, including the resistivity, magnetoresistance and Hall effect measurements, were carried out in commercial DynaCool Physical Properties Measurement System from Quantum Design. The resistivity and magnetoresistance were measured in a four-probe configuration and the Hall effect measurement was using a standard six-probe method.

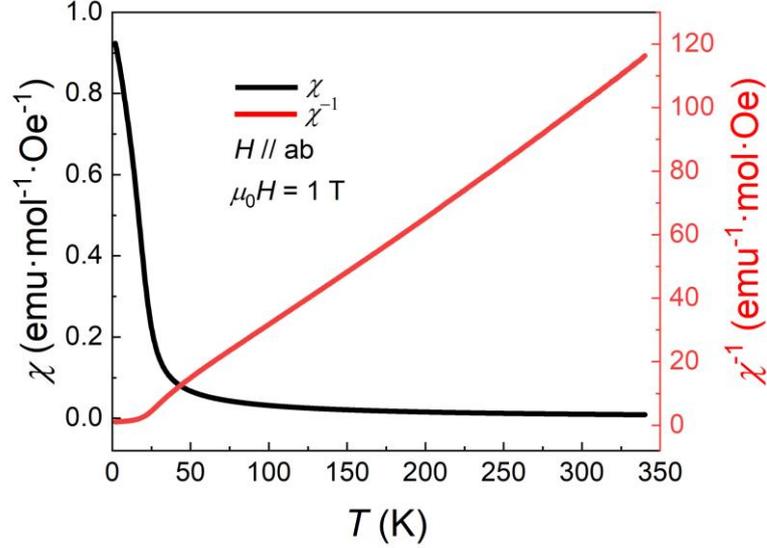

**Fig. S1.** The temperature dependent magnetic susceptibility $\chi$ and reciprocal magnetic susceptibility $\chi^{-1}$ under $\mu_0 H = 1$ T for $H \parallel ab$ plane.

The Curie-Weiss law was used to fit the reciprocal magnetic susceptibility $\chi^{-1}$ between 50 K to 350 K measured with $\mu_0 H = 1$ T and $H//ab$ plane, shown in Fig. S1, which gives the Weiss temperature $\theta_w = 11.09$ K and the effective moment $\mu_{eff} = 4.54$ $\mu_B$/Mn. The effective moment is close to the expected 5 $\mu_B$ for the $Mn^{2+}$ ion. The positive Weiss temperature indicates the FM ground state of $MnSb_6Te_{10}$, which is also supported by the clear hysteresis loop in $M(H)$ as shown in Fig. 1(d) of the main text.

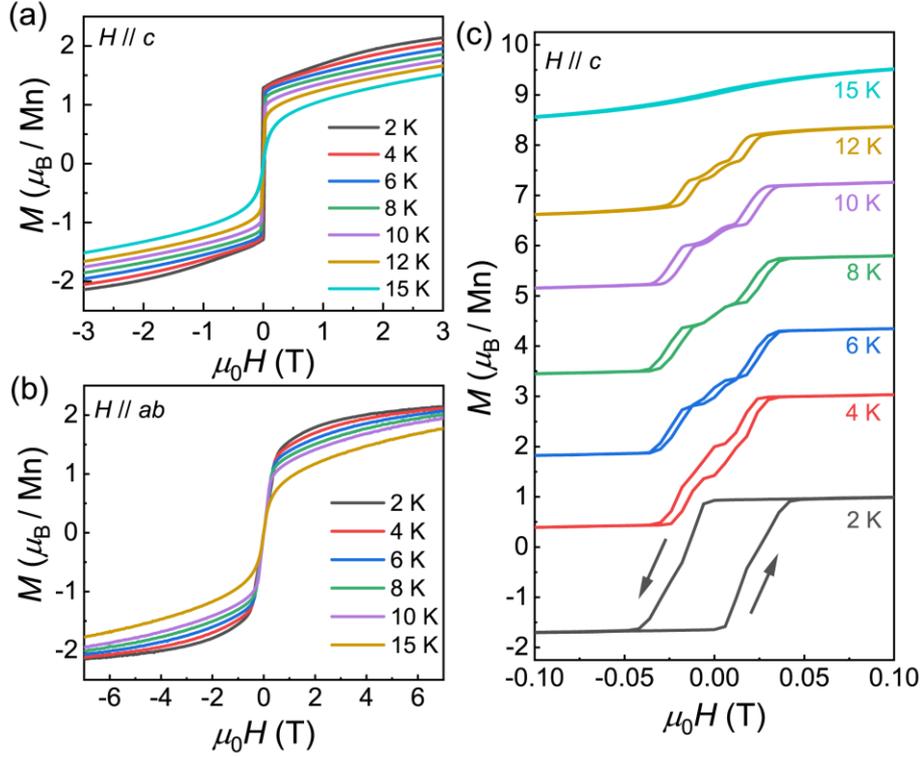

**Fig. S2.** (a)-(b) The isothermal magnetizations of MnSb$_6$Te$_{10}$ at various temperatures with $H//c$ axis and $H//ab$ plane, respectively. (c) The hysteresis loop of isothermal magnetizations at various temperatures with $H//c$ axis.

As mentioned in the main text, the long range AFM order could be identified in the magnetic susceptibility curves at $T_N$ = 12.7 K. When $H//c$, the ZFC and FC curves show clear separation at low temperature, indicating the possibility of hysteresis loops signifying the ferrimagnetic or FM state, demonstrated by the hysteresis loops of isothermal magnetizations as shown in Fig. S2(c), where a first-order spin-flop transition with hysteresis at 4 K beginning around $\mu_0H$ = 30 mT and vanishing at $\mu_0H$ = 40 mT is visible, indicating very soft magnetism. It is also clear that though the magnetic ground state is FM, it is fragile and becomes ferromagnetic or AFM at higher temperature. However, at $\mu_0H$ = 40 mT, the ferrimagnetic state also enters into the induced spin-polarized FM state. Comparing with the MnSb$_2$Te$_4$ and MnSb$_4$Te$_7$ [22, 23], the smaller saturation magnetic field of MnSb$_6$Te$_{10}$ indicates the weaker $Mn^{2+}$ - $Mn^{2+}$ interlayer AFM exchange. While increasing the temperature to 15 K, the hysteresis loop as well as the spin-flop transition gradually diminishes. The saturation moment at 2 K under $\mu_0H$ = 0.1 T is 1.7 $\mu_B$/$Mn^{2+}$, which is smaller than the value of 1.93$\mu_B$/$Mn^{2+}$ of $MnSb_4Te_7$ [23]. The remarkably reduced magnetic moment may due to that the Mn disorders and mixed location of Mn and Sb atoms [28], or is caused by the enhanced hole-carrier mediated Ruderman-Kittle-Kasuya-Yosida (RKKY) interaction which could give rise to magnetic frustration [29-31]. Figs. S2(a) and 2(b) display the isothermal magnetizations with *H//c* and *H//ab*, showing apparently that the saturation field for *H//c* is slightly smaller than that for *H//ab* at the same temperature, which indicates that the *c* axis is still the magnetic easy axis for $MnSb_6Te_{10}$.

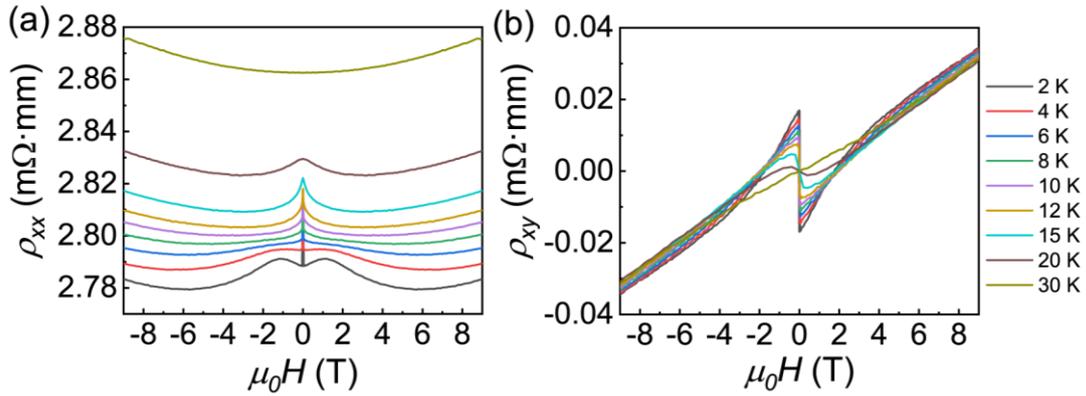

**Fig. S3.** Magnetotransport properties of bulk $MnSb_6Te_{10}$ single crystal. (a) Magnetoresistance $\rho_{xx}$ with *I//ab* plane and *H//c* axis at various temperatures. (b) Hall resistivity $\rho_{xy}$ measured at the same temperatures as in (a).

The magnetic field dependent $\rho_{xx}(H)$ and $\rho_{xy}(H)$, measured in the magnetic field range of $\mu_0H$ = -9 T - 9T are shown in Figs. S3(a) and 3(b), respectively. In Fig. S3(a), the $\rho_{xx}(H)$ curves exhibit an overall "M" shape, which is possibly caused by the combination of nonmagnetic parabolic MR and negative MR due to FM fluctuations. It is also clear that the M-shaped MR almost disappears and becomes a parabolic shape above $T_N$, suggesting the rather weak FM fluctuations above the $T_N$ in MnSb$_6$Te$_{10}$, which resembles the case of MnBi$_8$Te$_{13}$ [3]. As shown in Fig. S3(b), $\rho_{xy}(H)$ curves display double peaks feature at low magnetic field below $T_N$, likely signifying the topological Hall effect which is possibly related with the Berry curvature of the WSM state [32]. If it is true, the topological Hall effect could be viewed as an evidence for the FM Weyl state as we argued in the main text. The magnetotransport properties are tightly related with the magnetic structure. As displayed in Fig. S4, $\rho_{xx}(H)$ follows the same hysteresis loop as that in $M(H)$ with $H//c$ and $I//ab$ plane, where the red and black arrows represent the paths for changing the magnetic field. The clear hysteresis loop without the multiple transitions feature indicates the FM ground state of MnSb$_6$Te$_{10}$.

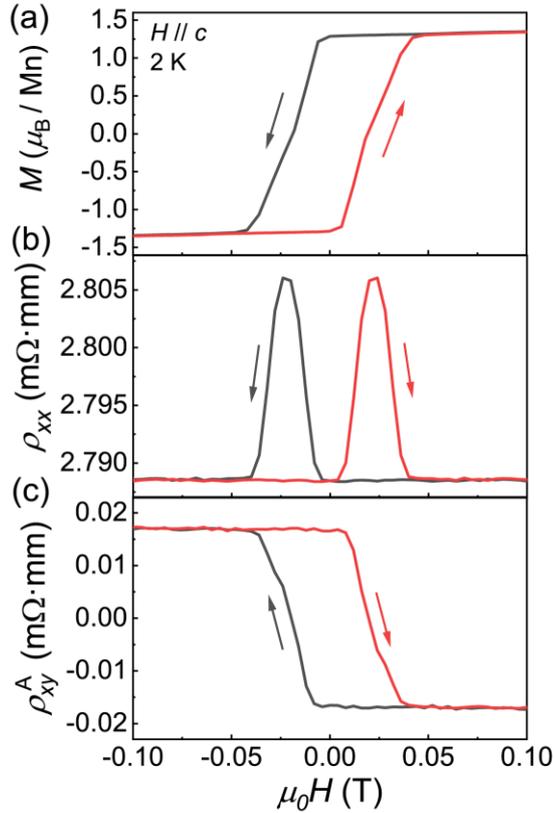

**Fig. S4.** Magnetotransport properties of bulk MnSb$_6$Te$_{10}$ single crystal. (a) The field dependence of magnetization *M*, (b) transverse magnetoresistance $\rho_{xx}$, and (c) anomalous Hall resistivity $\rho_{xy}^A$ at 2 K with *H*//*c* and *I*//*ab* plane.

## c. First principles calculations

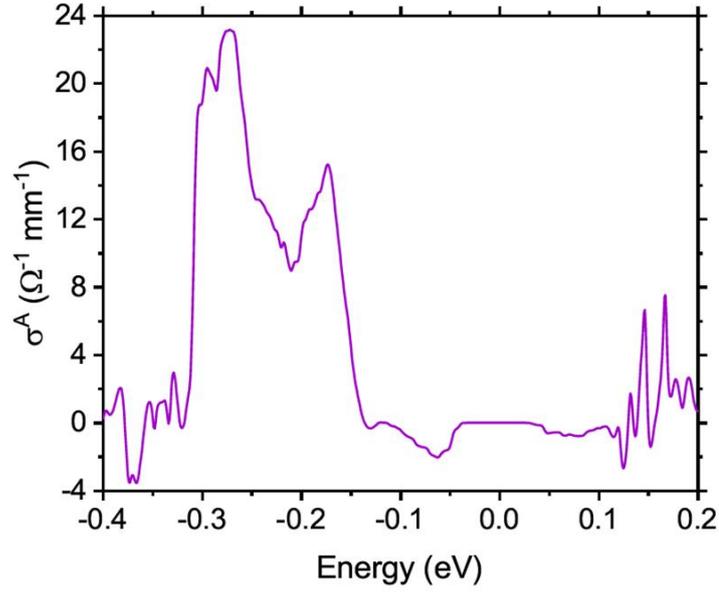

**Fig. S5.** The calculated anomalous Hall conductivities of MnSb$_6$Te$_{10}$ in the ferromagnetic FMz state. Here the Femi energy is set as zero.

We carried out first principles calculations on MnSb$_6$Te$_{10}$ in the framework of the generalized gradient approximation (GGA) functional [33] of the density functional theory through employing the Vienna *ab initio* simulation package (VASP) [34] with projector augmented wave method [35]. The on-site Hubbard U = 3 eV parameter on the Mn element are used in all calculations. The Bloch states are projected to the Wannier functions [36, 37] to build the tight-binding Hamiltonian, and then use irvsp program [38] [14] and WannierTools package [39] to calculate the anomalous Hall conductivity (AHC) with 151×151×101 k-point mesh as shown in Fig. S5. We can see that the calculated AHC at the -382 meV is 2.05 $\Omega^{-1}$ $mm^{-1}$ which is close to experimental value.

**Table S1.** The positions and chiralities of Weyl nodes near the Fermi level in the valence bands in the FMz phase. Here the bands are remarked by VB, VB-1, VB-2… from the highest valence band to lower valence band.

| Weyl nodes (ang$^{-1}$) | band | chirality |
| --- | --- | --- |
| (-0.182, 0.106, -0.013) | VB-2 and VB-3 | 1 |
| (0.182, -0.106, 0.013) | VB-2 and VB-3 | -1 |
| (0.000, -0.210, -0.013) | VB-2 and VB-3 | 1 |
| (0.000, 0.210, 0.013) | VB-2 and VB-3 | -1 |
| (-0.206, 0.119, 0.001) | VB-3 and VB-4 | -1 |
| (0.206, -0.119, -0.001) | VB-3 and VB-4 | 1 |

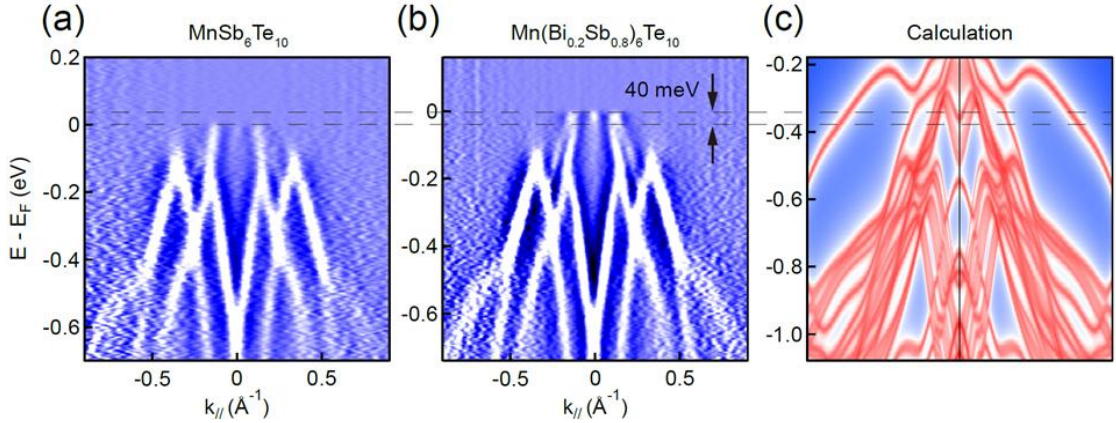

**Fig. S6.** ARPES measurements of pure and Bi doping MnSb$_6$Te$_{10}$ and Mn(Bi$_{0.2}$Sb$_{0.8}$)$_6$Te$_{10}$ single crystal. (a) ARPES results of pure MnSb$_6$Te$_{10}$ along K-Γ-K direction, (b) ARPES results of Mn(Bi$_{0.2}$Sb$_{0.8}$)$_6$Te$_{10}$ along K-Γ-K direction, and (c) DFT calculation of MnSb$_6$Te$_{10}$ along K-Γ-K direction.

### d.  ARPES measurements

ARPES measurements were performed at both 03U beamline of Shanghai Synchrotron Radiation Facility (SSRF) equipped with Scienta-Omicron DA30 electron analyzers. The angular and the energy resolutions were set to $0.2^o$ and $8 \sim 20$ meV (dependent on the selected probing photon energy). All samples were cleaved in an ultrahigh vacuum better than $8.0 \times 10^{-11}$ Torr. All the measurements are taken at 20 K.